\documentclass[pra,superscriptaddress,12pt,tightenlines,nofootinbib]{revtex4}

\usepackage{bm}
\input Qcircuit.tex

\newcommand{\tr}{{\rm tr}}

\newcommand{\vc}[2]%
{\left(\begin{array}{c}{\!\!#1\!\!}\\{\!\!#2\!\!}\end{array}\right)}

\begin{document}

\title{Subjective probability and quantum certainty}

\author{Carlton M.~Caves}
\email{caves@info.phys.unm.edu}
\affiliation{Department of Physics and Astronomy, MSC07--4220,
University of New Mexico, Albuquerque, NM~87131-0001, USA}
\author{Christopher A. Fuchs}
\email{cafuchs@research.bell-labs.com}
\affiliation{Bell Labs, Lucent
Technologies, 600--700 Mountain Avenue, Murray Hill, NJ 07974, USA}
\author{R\"udiger Schack}
\email{r.schack@rhul.ac.uk}
\affiliation{Department of Mathematics, Royal Holloway, University of
London, Egham, Surrey TW20$\;$0EX, UK}

\date{\today}

\begin{abstract}
In the Bayesian approach to quantum mechanics, probabilities---and thus
quantum states---represent an agent's degrees of belief, rather than
corresponding to objective properties of physical systems.  In this paper
we investigate the concept of {\em certainty\/} in quantum mechanics.
Particularly, we show how the probability-1 predictions derived from pure
quantum states highlight a fundamental difference between our Bayesian
approach, on the one hand, and Copenhagen and similar interpretations on
the other. We first review the main arguments for the general claim that
probabilities always represent degrees of belief. We then argue that a
quantum state prepared by some physical device always depends on an
agent's prior beliefs, implying that the probability-1 predictions derived
from that state also depend on the agent's prior beliefs. Quantum
certainty is therefore always some agent's certainty. Conversely, if facts
about an experimental setup could imply agent-independent certainty for a
measurement outcome, as in many Copenhagen-like interpretations, that
outcome would effectively correspond to a pre\"existing system property.
The idea that measurement outcomes occurring with certainty correspond to
pre\"existing system properties is, however, in conflict with locality.
We emphasize this by giving a version of an argument of Stairs [A.~Stairs,
Phil.\ Sci.\ {\bf 50}, 578 (1983)], which applies the Kochen-Specker
theorem to an entangled bipartite system.
\end{abstract}

\maketitle

\section{Introduction}
\label{sec:intro}

At the heart of Bayesian probability
theory~\cite{DeFinetti1990,Bernardo1994} is a strict category distinction
between propositions and probabilities.  Propositions are either true or
false; the truth value of a proposition is a fact.  A probability is an
agent's degree of belief about the truth of some proposition.  A
probability assignment is neither true nor false; probability assignments
are not propositions.

This category distinction carries over to probabilities for events or,
more particularly, for the outcomes of observations or measurements.  This
particular context, of measurements and their outcomes, is the most
relevant for this paper.  The proposition corresponding to an outcome of
an observation is the statement that the outcome occurs.  Ascertaining or
eliciting\footnote{We introduce ``eliciting'' here as an alternative to
``ascertaining'' because ``ascertaining'' has the connotation of
determining a pre\"existing property.  This, however, is in conflict with
the central point of our paper in the quantum-mechanical case.
 ``Eliciting'' at least suggests subtly that the outcome might have no
prior existence.} the outcome determines the truth value of the
proposition for the agent.  The outcome is thus a fact for the agent. He
can use the fact to modify his probabilities, but the probabilities
themselves are not facts.

Any actual usage of probability theory starts from an agent's prior
probability assignment.  Gathering data allows the agent to update his
probability assignments by using Bayes's rule. The updated probabilities
always depend on the agent's prior probabilities as well as on the data
and thus can be different for agents in possession of the same data.  It
is in this sense that probability assignments can be called {\em
subjective}, meaning they depend upon the agent.  For lack of a better
term, we adopt this usage in this paper.

Subjective probabilities are not arbitrary.  They acquire an operational
meaning in decision theory~\cite{Savage1972}. The Dutch-book
argument~\cite{DeFinetti1990} shows that, to avoid sure loss, an agent's
gambling commitments should obey the usual probability axioms. By
maintaining a strict category distinction between facts and probabilities,
Bayesian arguments make an explicit distinction between the objective and
subjective parts of any application of probability theory.  The {\em
subjective\/} part of a statistical argument is the initial judgment that
leads to prior probability assignments.  The {\em objective\/} part is any
given data and the application of the rules of probability theory to the
(subjective) prior probabilities.  This part is objective because neither
the data nor the rules of probability theory depend upon the agent's
beliefs.

Bayesian theory is conceptually straightforward. It provides simple and
compelling accounts of the analysis of repeated trials in
science~\cite{Savage1972}, statistical mechanics and
thermodynamics~\cite{Jaynes1957a,Jaynes1957b}, and general statistical
practice~\cite{Bernardo1994}.  Still, it might seem to have a limited
purview. For instance, a subjectivist interpretation of probability is
natural in a deterministic world, where the outcome of any observation can
be predicted with certainty given sufficient initial information.
Probabilities then simply reflect an agent's ignorance.  But what of an
indeterministic world?

In quantum mechanics the usual perception is that not all probabilities
can be interpreted as subjective degrees of
belief~\cite{Giere1973,Suppes1973,Giere1979,Loewer2001,Loewer2004,Ismael2006}.
Particularly, the probabilities of the outcomes of a quantum measurement
on a system in a pure quantum state are given by physical law and are
therefore objective---i.e., not depending on any agent's belief.  Or so
goes the usual argument.  We have shown in a series of previous
publications~\cite{Brun2001a,Fuchs2002,Caves2002a,Caves2002c,Caves2002b,Fuchs2004a,Schack2004,Caves2005a}
that, despite this common perception, all probabilities in quantum
mechanics can be interpreted as Bayesian degrees of belief and that the
Bayesian approach leads to a simple and consistent picture, which resolves
several of the conceptual difficulties of the interpretation of quantum
mechanics.  A consequence of the Bayesian approach is that all quantum
states, even pure states, must be regarded as subjective.

This is not to say, however, that everything in the quantum formalism is
subjective.  For instance, the Born rule for calculating probabilities
from quantum states is not subjective.  Instead it is akin to the rules of
probability theory itself, which, as pointed out above, make no reference
to an agent's particular beliefs.  We will return to this point and make
more of it in the Conclusion.

In this paper, our main aim is to address more carefully than previously
the problem of {\em certainty\/} in the Bayesian approach to quantum
mechanics. We show that a consistent treatment of quantum probabilities
requires that even if a measurement outcome has probability~1, implying
certainty about the outcome,\footnote{In this paper, we make no
distinction between probability 1 and certainty.  This is unproblematic
because we only consider measurements with a finite or countably infinite
number of outcomes.} that probability has to be interpreted as a Bayesian
degree of belief.  This is the case if the premeasurement state is an
eigenstate of the measured observable.  Even in this case we maintain, as
we must if we are to hold that pure states are subjective,\footnote{In a
previous publication~\cite{Caves2002c}, the authors were confused about
the status of certainty and pure-state assignments in quantum mechanics
and thus made statements about state preparation that we would now regard
as misleading or even wrong.  For much discussion surrounding this point
and sidelights on the material in the present paper, see C.~A. Fuchs, {\sl
Quantum States:\ What the Hell Are They?} (particularly, the essays on
pp.~35--113), available at {\tt
http://netlib.bell-labs.com/who/cafuchs/PhaseTransition.pdf}.} that the
measurement has no preassigned outcome.  There is no element of reality
that guarantees the particular measurement outcome.\footnote{This
distinction of ours can also be reworked in terms drawn from the
philosophy of language, as has been done by Timpson~\cite{Timpson2006}. It
contrasts with what is sometimes called the ``eigenstate-eigenvalue
link,'' which is almost universally adopted (in the strong form quoted
below) in the quantum foundations literature, even by treatments otherwise
sympathetic to a Bayesian view of quantum states.  For instance, Brukner
and Zeilinger~\cite{Brukner2001} give this description of the case of
certainty:
\begin{quote}
Only in the exceptional case of the qubit in an eigenstate of the
measurement apparatus the bit value observed reveals a property
already carried by the qubit.  Yet in general the value obtained by
the measurement has an element of irreducible randomness and
therefore cannot be assumed to reveal the bit value or even a hidden
property of the system existing before the measurement is performed.
\end{quote}}  Certainty is a function of the agent, not of the
system.\footnote{More philosophical, quantum-independent precedents
for this notion of `certainty' can be found in the discussion given
by White~\cite{WhiteWittgenstein}. White makes a distinction between
the certainty of agents and the certainty of things in themselves,
and puts it this way:
\begin{quote}
The certainty of persons and the certainty of things are logically
independent of each other.  Somebody can be (or feel) certain of
something which is not itself certain, while something can be certain
without anybody's being (or feeling) certain of it.  `He is certain
that $p$' neither implies nor is implied by the impersonal `It is
certain that $p$'.  The same thing cannot be both certain and not
certain, though one person can be certain of it and another of its
opposite ...  People can become more or less certain of something
which itself has not become any more or less certain.  ... A gambler
need not feel that {\it it\/} [our italics, for White is speaking of
the thing in this case] is certain that red will turn up next in
order to feel certain that it will.
\end{quote}}

Along with this, we give a precise account of a fundamental
difference between our Bayesian approach, on the one hand, and
various Copenhagen-like interpretations of quantum mechanics on the
other.\footnote{There exist many alternative attempts to solve the
conceptual problems of quantum theory. Most notable are the
many-worlds (or Everett) interpretation, hidden-variable theories
(e.g., Bohmian mechanics), and spontaneous-collapse theories (e.g.,
the Ghirardi-Rimini-Weber model). All of these are the subjects of
ongoing debates. In this paper, we do not comment on any of these
alternatives.} There are, of course, many versions of the Copenhagen
interpretation; here we focus on realist readings of
it~\cite{Faye1994,Murdoch1987,Rosenfeld1979,Jammer1974} (as opposed
to anti-realist readings~\cite{Faye1991,Plotnitsky1994} and other
more subtle interpretations~\cite{Folse1985}), which are the most
predominant in the physics community. To our knowledge, these all
have in common that a system's quantum state is determined by a
sufficiently detailed, agent-independent {\it classical
description\/} of the preparation device, which is itself thought of
as an agent-independent physical system~\cite{Peres1984,Stapp1972}.
This is the only salient feature of the interpretations we consider
here.  Although this feature is often associated with Copenhagen-like
interpretations,\footnote{We offer the following quotes as a small
justification for associating the objective-preparations view with
Copenhagen-like interpretations and, specifically, with Bohr.  The
first is drawn from a letter from Niels Bohr to Wolfgang Pauli, 2
March 1955 (provided to us by H.~J. Folse):
\begin{quote}
In all unambiguous account it is indeed a primary demand that the
separation between the observing subject and the objective content of
communication is clearly defined and agreed upon. \dots\ Of course, one
might say that the trend of modern physics is the attention to the
observational problem and that just in this respect a way is bridged
between physics and other fields of human knowledge and interest. But it
appears that what we have really learned in physics is how to eliminate
subjective elements in the account of experience, and it is rather this
recognition which in turn offers guidance as regards objective description
in other fields of science. To my mind, this situation is well described
by the phrase `detached observer' .... Just as Einstein himself has shown
how in relativity theory `the ideal of the detached observer' can be
retained by emphasizing that coincidences of events are common to all
observers, we have in quantum physics attained the same goal by
recognizing that we are always speaking of well defined observations
obtained under specified experimental conditions. These conditions can be
communicated to everyone who also can convince himself of the factual
character of the observations by looking on the permanent marks on the
photographic plates. In this respect, it makes no difference that in
quantum physics the relationship between the experimental conditions and
the observations are of a more general type than in classical physics.
\end{quote}
The second comes from Bohr in Ref.~\cite{Bohr1939}:
\begin{quote}
In the system to which the quantum mechanical formalism is applied,
it is of course possible to include any intermediate auxiliary agency
employed in the measuring process [but] some ultimate measuring
instruments must always be described entirely on classical lines, and
consequently kept outside the system subject to quantum mechanical
treatment.
\end{quote}}
we prefer to use the neutral term ``objective-preparations view'' to
refer to it throughout the remainder of this paper.  The most
important consequence of this feature is that, with it, quantum
states must be regarded as objective.\footnote{Even more extreme is
this reading of Bohr, from p.~107 of Ref.~\cite{Murdoch1987}:
\begin{quote}
Bohr, then, held what I shall call the {\it objective-values
theory\/} of measurement, according to which successful observation
or measurement reveals the objective, pre-existing value of an
observable. \ldots\ It is important here to note that the
objective-values theory should not be confused with what I shall call
the {\it intrinsic-values theory\/} of properties \ldots\ according
to which all the observables of an object have, at any moment,
definite values.  The latter theory, which Bohr rejected, is
logically independent of the former.
\end{quote}}  Measurement outcomes that have probability~1 bring
this difference into stark relief.  In the objective-preparations view, a
probability-1 outcome is {\it objectively certain}, guaranteed by facts
about the preparation device, and thus corresponds to a pre\"existing
property of the agent's external world.

The paper is organized as follows. In Sec.~\ref{sec:obj}, we discuss
the strict category distinction between facts and probabilities
within the setting of applications of probability theory to classical
systems.  We argue, following de Finetti~\cite{DeFinetti1931}, that
in the last analysis probability assignments are always subjective in
the sense defined earlier. We briefly consider the concept of
objective probability, or {\em objective chance}, and review the main
argument showing that this concept is problematic within the
classical setting.

Section~\ref{sec:prior} addresses the role of prior belief in quantum
state preparation.  In our interpretation, the quantum state of a system
cannot be determined by facts alone.  In addition to the facts an agent
acquires about the preparation procedure, his quantum state assignment
inevitably depends on his prior beliefs.  This constitutes the central
difference between the Bayesian approach to quantum mechanics and the
objective-preparations view.  We argue that, by positing that states are
fully determined by facts alone, the objective-preparations view neglects
to take into account that quantum mechanics applies to preparation
devices, even when derived from the ``ultimate measuring instruments'' of
Ref.~\cite{Bohr1939}.

In Sec.~\ref{sec:penr} we turn to the question of certainty in quantum
mechanics and emphasize that in the objective-preparations view, facts
about an experimental setup imply objective, agent-independent certainty
for the outcomes of appropriate measurements.  The objective-preparations
view thus implies a pre\"existing property of the agent's external world
guaranteeing the measurement outcome in question---a point we find
untenable. In Sec.~\ref{sec:koch}, we support the view that measurement
outcomes occurring with certainty cannot correspond to pre\"existing
properties by showing it to be in conflict with locality. For this
purpose, we use a modification of an argument of Stairs~\cite{Stairs1983}
(independent variations of the argument can be found
in~\cite{Heywood1983} and~\cite{Brown1990}), which applies the
Kochen-Specker noncolorability theorem to an entangled bipartite system.

Finally, in Sec.~\ref{sec:cert} we give a short general discussion on the
meaning of certainty in a world without pre\"existing instruction sets for
quantum measurement outcomes~\cite{Mermin1985}.  We emphasize that
certainty is always an agent's certainty; there is nothing in the physical
world itself that makes a quantum-mechanical probability-1 prediction true
before the act of finding a measurement outcome.  We conclude with brief
discussions of the status of the Born rule and directions for further work
on the Bayesian approach.

\section{Subjective probability versus objective chance}
\label{sec:obj}

The starting point for our considerations is a category distinction.
Probability theory has two main ingredients.  Firstly, there are {\em
events}, or {\em propositions}.  Mathematically, the space of events forms
a sigma algebra.  Conceptually, what is important about events is that, at
least in principle, an agent can unambiguously determine whether an event
has occurred or not.  Expressed in terms of propositions, the criterion is
that the agent can determine unambiguously in what circumstances he would
call a proposition true and in what others false~\cite{DeFinetti1990}. The
occurrence or nonoccurrence of an event is a {\em fact\/} for the agent.
Similarly, the truth or falsehood of a proposition is a fact. We shall say
that facts are {\em objective}, because they are not functions of the
agent's beliefs.

Secondly, there are {\em probabilities}. Mathematically, probabilities are
measures on the space of events. Probabilities are fundamentally different
from propositions. Theorems of probability theory take an event space and
a probability measure as their starting point. Any usage of probability
theory starts from a {\em prior probability assignment}.  The question of
whether a prior probability assignment is true or false cannot be
answered.

The subjectivist Bayesian approach to probability
theory~\cite{DeFinetti1990,Savage1972,Bernardo1994,Jeffrey2002a,Kyburg1980}
takes this category distinction as its foundation.  Probabilities are
degrees of belief, not facts. Probabilities cannot be derived from facts
alone. Two agents who agree on the facts can legitimately assign different
prior probabilities. In this sense, probabilities are not objective, but
\hbox{\em subjective}.

Even though we hold that {\it all\/} probabilities, classical and quantum,
are Bayesian, the reader is encouraged to view this section as
predominantly a discussion of applications of probability theory to
classical systems and the subsequent sections as having to do mainly with
quantum probabilities. In both settings, when we refer to facts, we are
usually thinking about data, in the form of outcomes or results, gathered
from observations or measurements.

It is often said that one can verify a probability, thereby making it a
fact, by performing repeated trials on independent, identically
distributed systems. What is missed in this statement is that the repeated
trials involve a bigger event space, the space of all potential sequences
of outcomes.  To apply probability theory to repeated trials requires
assigning probabilities to these potential sequences, and additional
subjective judgments are required to do this. (For two very nice
expositions of this point, see~\cite{Appleby2005a,Appleby2005b}.)
Moreover, the outcome frequencies observed in repeated trials are not
probabilities.  They are facts.  Like any fact, an observed frequency can
be used, through Bayes's rule, to update subjective probabilities, in this
case the subjective probabilities for subsequent trials.

Consider tossing a coin.  The probabilities for Heads and Tails cannot be
derived from physical properties of the coin or its environment.  To say,
for example, that the coin is ``fair'' is ultimately a subjective
judgment, equivalent to assigning a symmetric prior probability to the
possible outcomes.  A symmetry argument applied to facts of the mass
distribution of the coin does not determine probabilities, because the
outcome of a toss also depends on the initial conditions.  These, too, are
facts, but a symmetry argument applied to the initial conditions must be
phrased in terms of probabilities for the initial conditions, i.e., in
terms of judgments, not facts.  In coming to the judgment that a coin is a
fair coin, an agent is well advised, of course, to take into account all
known facts about the physical constitution of the coin, perhaps even
developing a detailed model of the mass distribution of the coin and the
coin-tossing mechanism. In that case, however, the agent must still make
probabilistic judgments at an earlier stage of the model, say, regarding
the initial conditions for the tossing mechanism or the state of the
surrounding gas. Probability assignments are not arbitrary, but they
always have an irreducibly subjective component.

To summarize, even probabilities that follow from symmetry arguments are
subjective, because the symmetry argument is applied to the probabilities,
not to facts.  The assumed symmetry is an agent's judgment about the
events in question, and the resulting probabilities are the expression of
that judgment.  This paper takes one important further step: Our central
claim is that even probabilities that appear to be given by physical law,
as in quantum theory, are subjective.

The Bayesian approach is immediately applicable to physics experiments
because it accounts effortlessly for repeated trials~\cite{Caves2002b}.
For an excellent general discussion of the use of subjective probability
in science, see Chapter~4 of~\cite{Savage1972}. An example of an area
where subjective probabilities have had notable success is classical
statistical mechanics~\cite{Jaynes1957a,Jaynes1957b}, where the Bayesian
approach draws a strict category distinction between a system's
microstate, which is a fact, and the subjective probabilities assigned to
microstates, which are a reflection of an agent's ignorance of the
microstate.  Even here, however, the success is often belittled because of
a failure to appreciate the category
distinction~\cite{Shalizi-0303,North2003}. Suppose an agent assigns an
epistemic uniform probability distribution to an ice cube's microstates.
The ice cube melts. Does the ice cube melt, it is asked, because of the
agent's probability assignment~\cite{North2003}?  How can an agent's
epistemic state have anything whatsoever to do with the ice's melting?
The answer to these questions is simple: An agent's epistemic state is
part of the reason for his prediction that the ice cube melts, not part of
the reason for the melting.  The ice will either melt or not melt; it is
indifferent to the agent's ignorance. ``This ice cube will melt'' is a
proposition whose truth value is a fact about the world. ``Ice melts'' is
an abbreviated version of the subjective judgment that the probability is
close to 1 that a typical ice cube will melt.  If an agent were able to
determine the ice cube's initial microstate, this would have no effect on
whether the ice melts, but it would mean that the agent could extract more
energy from the process than somebody else who is ignorant of the
microstate.

Despite the successes of the Bayesian approach to probability in physics,
there appears to be a strong desire, among a sizeable number of
physicists, for an objective probability concept. We now review the main
argument against the validity of such a concept within the classical
setting.

In physics, probabilities appear side by side with physical parameters
such as length and are used in a superficially similar way. Both length
and probability appear in mathematical expressions used for predicting
measurement outcomes. This has led to attempts, for instance by
Braithwaite~\cite{Braithwaite1968}, to treat probability statements as
physical parameters residing in the category of facts. A probability
statement such as ``the probability of this atom decaying in the next 5
minutes is $p=0.3$'' would thus be a proposition, analogous to, e.g., a
geometrical statement such as ``the length of this ruler is $l=0.3\,$m''.
This geometrical analogy~\cite{Feller1968} is inherently flawed,
however~\cite{DeFinetti1931}.  Whereas the truth of a statement concerning
the length of a ruler can be unambiguously decided (at least in the
approximate form $0.29\mbox{m}\le l\le0.31\mbox{m}$), the truth of
statements concerning probabilities cannot be decided, not even
approximately, and not even in principle.  The usual method of
``verifying'' probabilities, through the outcomes of repeated trials,
yields outcome frequencies, which belong to the category of events and
propositions and are not probabilities.  Probability theory allows one to
assign a probability (e.g., $p=0.99$) to the proposition ``the outcome
frequency is in the interval $0.29\le f\le0.31$'', and the truth of this
proposition can be unambiguously decided, but this is a proposition about
an outcome frequency, not about probability.

In order to bridge the category distinction between probability and
physical parameters, a new principle or axiom is needed. In Braithwaite's
theory, for instance, the new principle is introduced in the form of the
acceptance and rejection rules of orthodox statistics~\cite{Lehmann1986}.
These rules must be postulated, however, because they cannot be
systematically derived from probability theory~\cite{Braithwaite1968}.
Even though most of non-Bayesian statistics is based on these rules, they
are essentially {\em ad hoc}, and the way they are used in statistics is
highly problematical~\cite{Jeffreys1961,Berger1987}.

Instead of discussing the conceptual difficulties of orthodox
statistics, we focus on a simpler and more direct postulate designed
to bridge the category distinction between probability and physical
parameters, namely, Lewis's Principal
Principle~\cite{Lewis1986a,Lewis1986b}. The Principal Principle (PP)
distinguishes between {\em chance\/} and Bayesian probability. Chance
is supposed to be objective.  The numerical value of chance is a
fact. Chance therefore belongs to the same category as a physical
parameter. If $E$ is an event, and $0\le q\le1$, the statement ``the
chance of $E$ is $q$'' is a proposition. Denote this proposition by
$C$. The Principal Principle links chance and probability by
requiring that an agent's conditional probability of $E$, given $C$,
must be $q$, irrespective of any observed data.  More precisely, if
$D$ refers to some other compatible event, e.g., frequency data, then
the Principal Principle states that the Bayesian probability must
satisfy
\begin{equation}
\label{eq:pp}
\Pr(E|C\&D) = q \;.
\end{equation}

Within the context of experimental situations with large sample sizes,
where Bayesian updating leads to similar posteriors for exchangeable
priors, the geometric analogy, combined with the Principal Principle to
connect chance with probability, would appear to work quite well.  This
gives rise to the idea that the Principal Principle accounts for the
concept of objective chance in physics. However, from a Bayesian
perspective, the introduction of chance is completely
unmotivated~\cite{DeFinetti1931,Jeffrey2002a,Jeffrey1997}. More urgently,
in those cases where the idea is not already fraught with obvious
difficulties, it serves no role that Bayesian probability itself cannot
handle~\cite{Jeffrey2002a}.

To illustrate one such difficulty, return to the coin-tossing example
discussed above, and assume that there is an objective chance $q$ that a
coin-tossing event will produce Heads.  As we have seen in the discussion
above, the chance cannot be deduced from physical properties of the coin
alone, because the probability of Heads also depends on initial conditions
and perhaps other factors.  An advocate of objective chance is forced to
say that the chance is a property of the entire ``chance situation,''
including the initial conditions and any other relevant factors.  Yet a
sufficiently precise specification of these factors would determine the
outcome, leaving no chance at all.  The circumstances of successive tosses
must be different to give rise to chance, but if chance aspires to
objectivity, the circumstances must also be the same.  Different, but the
same---there is no way out of this conundrum as long as objective and
chance are forced to co-exist in a single phrase.  Subjective
probabilities easily dispense with this conundrum by maintaining the
category distinction.  The differences between successive trials are
differences in the objective facts of the initial conditions; the sameness
is an agent's judgment that he cannot discern the differences in initial
conditions and thus assigns the same probability to every trial.

But what of probabilities in quantum mechanics?  Given the last paragraph,
one might well think---and many have thought---there is something
different going on in the quantum case.  For, in repeating a preparation
of a pure state $|\psi\rangle$, aren't all the conditions of preparation
the same by definition?  Any subsequent probabilities for measurement
outcomes will then be determined by applying the Born rule to
$|\psi\rangle$.  They are not subjective probabilities that come about by
an inability to take all circumstances into account.  Thus quantum states
(and hence quantum ``chances'') are objective after all, and the Principal
Principle is just the kind of thing needed to connect these quantum
chances to an agent's subjective probabilities---or so a very beguiling
account might run.

The objective-preparations view supports the seeming need for a PP-style
account by positing that classical facts about a preparation device
determine the prepared quantum state and its associated measurement
probabilities.  The subjective Bayesian interpretation of quantum
probabilities contends, in contrast, that facts alone never determine a
quantum state.  What the objective-preparations view leaves out is the
{\it essential\/} quantum nature of the preparation device, which means
that the prepared quantum state always depends on prior beliefs in the
guise of a quantum operation that describes the preparation device.  We
turn now to a discussion of these issues in the next two sections.

\section{Prior beliefs in quantum state preparation}
\label{sec:prior}

In the subjectivist interpretation of quantum-mechanical probabilities
advocated in this paper, the strict category distinction between
(objective) facts and (subjective) probabilities holds for all
probabilities, including probabilities for the outcomes (facts) of quantum
measurements. Since probabilities are an agent's subjective degrees of
belief about the possible outcomes of a trial and quantum states are
catalogues of probabilities for measurement outcomes, it follows that
quantum states summarize an agent's degrees of belief about the potential
outcomes of quantum measurements.  This approach underlines the central
role of the agent, or observer, in the very formulation of quantum
mechanics.  In this sense our interpretation is close to Copenhagen-like
interpretations, even when these interpretations incorporate the
objective-preparations view, but the Bayesian approach differs markedly
from the objective-preparations view in the way facts and quantum states
are related.

In the objective-preparations view, the facts about a classical
preparation procedure determine the quantum
state~\cite{Peres1984,Stapp1972}. According to the objective-preparations
view, one can give, in unambiguous terms, a description of an experimental
device that prepares a given quantum state; thus a quantum state is
completely determined by the preparation procedure. In the
objective-preparations view, there is no room for prior beliefs in quantum
state preparation; quantum states and the probabilities derived from them
are determined by objective facts about the preparation device.

In our interpretation, the quantum state of a system is not determined by
classical facts alone.  In addition to the facts, an agent's quantum state
assignment depends on his prior beliefs. We now show why this must be so.

Classically, Bayes's rule,
\begin{equation}
\Pr(h|d) = \frac{ \Pr(d|h) \Pr(h) }{ \Pr(d) }\;,
\end{equation}
is used to update probabilities for hypotheses $h$ after acquiring facts
in the form of data~$d$.  The posterior probability, $\Pr(h|d)$, depends
on the observed data $d$ and on prior beliefs through the prior
probabilities $\Pr(h)$ and the conditional probabilities
$\Pr(d|h)$.\footnote{Bayesian updating is consistent, as it should be,
with logical deduction of facts from other facts, as when the observed
data $d$ logically imply a particular hypothesis $h_0$, i.e., when
$\Pr(d|h)=0$ for $h\ne h_0$, thus making $\Pr(h_0|d)=1$.  Since the
authors disagree on the implications of this consistency, it is fortunate
that it is irrelevant to the point of this paper.  That point concerns the
status of quantum measurement outcomes and their probabilities, and
quantum measurement outcomes are not related by logical implication.  Thus
we do not discuss further this consistency, or its implications or lack
thereof.}

In quantum mechanics, the most general updating rule has the form
\begin{equation}
\rho \mapsto \rho_d  = \frac{{\cal A}_d(\rho)}{p_d} \;.
\label{kerplunk}
\end{equation}
Here $d$ is an observed measurement outcome (a fact); $\rho_d$ is the
post-measurement (posterior) state;  $\rho$ is the premeasurement (prior)
state; and ${\cal A}_d$ is a completely positive linear map, called a {\em
quantum operation}, corresponding to outcome~$d$ and given by
\begin{equation}    \label{eq:cpm}
{\cal A}_d(\rho)=\sum_j A_{dj}\rho A^\dagger_{dj}\;.
\end{equation}
The linear operators $A_{dj}$ define POVM elements
\begin{equation}
E_d = \sum_{j} A^\dagger_{dj} A_{dj}\;,
\end{equation}
which obey the normalization condition
\begin{equation}
\sum_{d} E_d = 1\;,
\end{equation}
and
\begin{equation}
p_d(\rho) = \tr \rho E_d = \tr\sum_j A_{dj}\rho A^\dagger_{dj}
\end{equation}
is the probability for outcome~$d$.  Similar to the classical case, the
posterior state depends on the measurement outcome, the prior state, and
the completely positive map ${\cal A}_d$, which is analogous to the
conditional probabilities of Bayesian
updating~\cite{Fuchs2002,Leifer2006a,Leifer2006b}.

We now argue that the posterior state always depends on prior beliefs,
even in the case of {\em quantum state preparation}, which is the special
case of quantum updating in which the posterior state is independent of
the prior state.  To be precise, in the state-preparation case, there is a
state $\sigma$ such that for outcome~$d$,
\begin{equation}
{\cal A}_d(\rho) = \sum_j A_{dj} \rho A^\dagger_{dj} = p_d(\rho)\,\sigma
\end{equation}
for all states $\rho$ for which $p_d(\rho)\ne 0$.  If the preparation
depends on obtaining a particular measurement outcome $d$, i.e., if
$p_d(\rho)<1$ for some $\rho$, the preparation operation is called {\it
stochastic\/}; if $p_d(\rho)=1$ for all $\rho$, the preparation device is
{\it deterministic}.  Notice that if the posterior state $\sigma$ is a
pure state $\ket\psi$, it corresponds to certainty for the outcome of a
yes-no measurement of the observable $O=\ket\psi\bra\psi$.

It is tempting to conclude that objective facts, consisting of the
measurement outcome~$d$ and a classical description of the preparation
device, determine the prepared quantum state~$\sigma$. This would violate
the category distinction by allowing facts to fully determine
probabilities derived from $\sigma$. What this can only mean for a
thoroughgoing Bayesian interpretation of quantum probabilities is that the
posterior quantum state $\sigma$ {\it must\/} depend on prior beliefs
through the quantum operation~\cite{Fuchs2002,Fuchs2004a,Fuchs2004b}. We
now consider this crucial difference in more detail.

The quantum operation depends, at least partly, on an agent's beliefs
about the device that executes the state-preparation procedure.  Any
attempt to give a complete specification of the preparation device in
terms of classical facts (i.e., observations or measurements of the device
and its method of operating) and thus to derive the quantum operation from
classical facts alone comes up against the device's quantum-mechanical
nature.

\begin{figure*}[t]
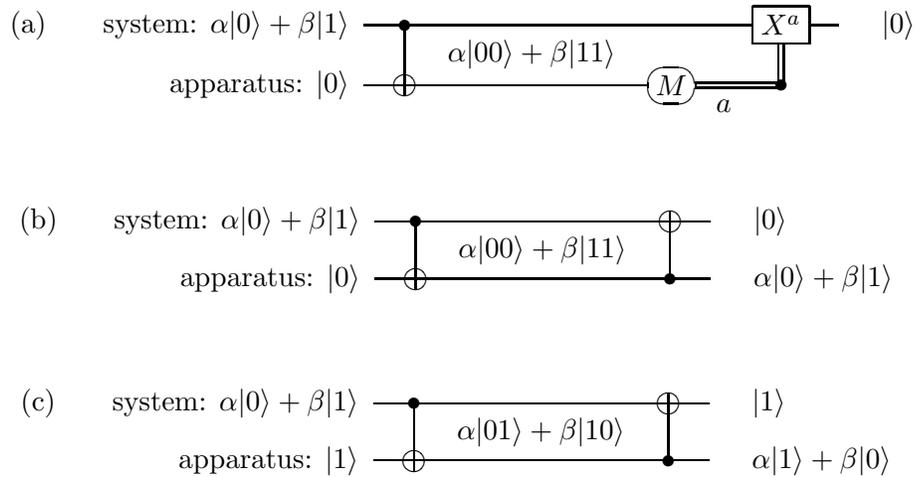

\[
\mbox{(a)}\hspace{120pt}
\begin{Qcircuit} @C=1em @R=0.8em @!R{
\lstick{\mbox{system: $\alpha\ket0+\beta\ket1$}}&\ctrl{1}&\qw&\qw&\qw&\dstick{\alpha\ket{00}+\beta\ket{11}}\qw&\qw&\qw&\qw&\qw        &\qw          &\gate{X^a}     &\qw&\rstick{\ket0} \\
\lstick{\mbox{apparatus: $\ket0$}}              &\targ   &\qw&\qw&\qw&\qw                                     &\qw&\qw&\qw&\measure{M}&\dstick{a}\cw&\control\cw\cwx&
}
\end{Qcircuit}
\]
\vspace{24pt}
\[
\hspace{-45pt}\mbox{(b)}\hspace{120pt}
\begin{Qcircuit} @C=1em @R=1.3em @!R{
\lstick{\mbox{system: $\alpha\ket0+\beta\ket1$}}&\ctrl{1}&\qw&\qw&\qw&\dstick{\alpha\ket{00}+\beta\ket{11}}\qw&\qw&\qw&\qw&\targ    &\qw&\rstick{\ket0} \\
\lstick{\mbox{apparatus: $\ket0$}}              &\targ   &\qw&\qw&\qw&\qw                                     &\qw&\qw&\qw&\ctrl{-1}&\qw&\rstick{\alpha\ket0+\beta\ket1}
}
\end{Qcircuit}
\]
\vspace{24pt}
\[
\hspace{-45pt}\mbox{(c)}\hspace{120pt}
\begin{Qcircuit} @C=1em @R=1.3em @!R{
\lstick{\mbox{system: $\alpha\ket0+\beta\ket1$}}&\ctrl{1}&\qw&\qw&\qw&\dstick{\alpha\ket{01}+\beta\ket{10}}\qw&\qw&\qw&\qw&\targ    &\qw&\rstick{\ket1} \\
\lstick{\mbox{apparatus: $\ket1$}}              &\targ   &\qw&\qw&\qw&\qw                                     &\qw&\qw&\qw&\ctrl{-1}&\qw&\rstick{\alpha\ket1+\beta\ket0}
}
\end{Qcircuit}
\vspace{12pt}
\]
\caption{\footnotesize\baselineskip=12pt (a)~Quantum-circuit diagram for a
device that prepares the system qubit in the state $\ket0$: The
controlled-NOT gate puts the system qubit and the apparatus qubit in the
entangled state $\alpha\ket{00}+\beta\ket{11}$; a measurement of the
apparatus qubit yields result $a=0$ or 1; the system state is flipped if
$a=1$, thus always preparing the system qubit in state $|0\rangle$,
regardless of the system's initial state.  The single lines in the circuit
diagram carry quantum states, which are subjective in the Bayesian view,
and the double lines carry the outcome $a$ of the measurement, which is an
objective fact that is used to conditionally flip the system qubit.  This
quantum circuit describes passing a spin-$1/2$ particle (photon) through a
Stern-Gerlach magnetic field (polarizing beam splitter), which sends
spin-up (horizontal polarization) and spin-down (vertical polarization)
along different paths; measuring which path; and then flipping the spin
(rotating the polarization from vertical to horizontal) if the particle is
determined to be moving along the spin-down (vertical-polarization) path.
This is a deterministic preparation device. Any deterministic preparation
operation can be realized in this way: entangle system and apparatus,
measure apparatus, and then change the system state conditional on the
measurement outcome. (b)~In the circuit of~(a), the flip based on the
measurement result $a$ can be moved in front of the measurement, becoming
a controlled-NOT gate. The measurement can then be omitted, making the
preparation device into a unitary interaction between system and
apparatus.  Any deterministic preparation operation can be realized by
such a purely unitary interaction. (c)~If the initial apparatus state in
(b) is changed to $|1\rangle$, the system qubit is prepared in the state
$|1\rangle$. Indeed,~(b) and (c) show that the system qubit is prepared in
a state identical to the initial apparatus state, whatever that state is.
This highlights our conclusion that the operation of a preparation device
always depends on prior beliefs about the device, in particular, its
initial quantum state.  The objective-preparations view, by positing that
the operation of a preparation device can be specified completely in terms
of facts, is forced to conclude that the input state to the apparatus is
an objective fact.  Thus if one adopts the objective-preparations view, one
is forced to regard quantum states as objective, a conclusion we reach by
a different route in Sec.~\ref{sec:penr}.}
\label{F:circuit}
\end{figure*}

Classical facts cannot suffice to specify a preparation device completely
because a complete description must ascribe to the device an initial
quantum state, which inevitably represents prior beliefs of the agent who
is attempting to describe the device.  It is quite possible---indeed,
likely---that other subjective judgments are involved in an agent's
description of the preparation device, but to find a prior belief that is
unavoidably part of the description, it is sufficient to recall the usual
justification for the mapping given in Eq.~(\ref{kerplunk})
\cite{Nielsen2000}.  The mapping can always be modeled as coming about
from a unitary interaction between the system and an apparatus, followed
by an observation on the apparatus alone.  To say what the unitary
operation actually does, however, one must specify initial quantum states
for all systems concerned.  But which quantum state for the apparatus?
That, the subjective Bayesian would say, is subjective.  The
objective-preparations assumption that a preparation device can be given a
complete classical description neglects that any such device is quantum
mechanical and thus cannot be specified completely in terms of classical
facts.  An example of the dependence of the system's output state on the
input state of the apparatus is given in Fig.~\ref{F:circuit}.
Mermin~\cite{Mermin2006} analyzes the same preparation apparatus, but
reaches quite different conclusions.

There is an analogy between the coin-tossing example discussed above and
the quantum-mechanical analysis of a preparation device.  By examining the
coin and the tossing mechanism, a scientist cannot derive the
probabilities for Heads and Tails.  These always depend on some prior
judgment.  Similarly, by examining a preparation device, a scientist
cannot derive the output quantum state and its associated probabilities
for measurement outcomes.  These also always depend on some prior
judgment.  The analogy cannot be pushed too far, however, even though in
both cases, facts never determine probabilities.  In the classical
setting, a complete specification of the coin's physical properties, the
tossing mechanism, and the initial conditions leads to certainty for the
outcome.  For a quantum preparation device, a complete specification of
the device in terms of facts is simply not allowed by the quantum
formalism.

In practice, of course, experimenters depend on their experience and on
manufacturers' specifications to inform the prior judgment.  They also use
repeated trials to {\em test\/} the entire setup~\cite{D'Ariano2004}.  To
analyze the test results, however, requires the use of the
quantum-mechanical formalism, which inevitably involves a prior judgment
as input.  An example of such a prior judgment is the assumption of
exchangeability for repeated trials; for thorough discussions of
exchangeability in quantum tomography of states and operations, the reader
is urged to consult our previous papers on these
subjects~\cite{Caves2002b,Fuchs2004a,Fuchs2004b}.

An important consequence of our argument is that a quantum operation
assigned to a preparation device belongs to the same category in our
category distinction as quantum states.  This must be so because such a
quantum operation determines its output state irrespective of the input.
The quantum operations assigned to preparation devices are therefore
subjective.  The subjectivity of these quantum operations is akin to the
subjectivity of conditional probabilities in probability theory.

It follows from all this that two agents who agree on all the facts
relevant to a quantum experiment can disagree on the state assignments. In
general, two agents starting from the same facts, but different priors,
arrive at different (posterior) state assignments. For sufficiently
divergent priors, the two agents might even legitimately assign different
pure states~\cite{Caves2002a}, as in the example of Fig.~\ref{F:circuit}.

\section{Certainty and objective properties}
\label{sec:penr}

The previous section can be summarized as follows. A crucially important
difference between the objective-preparations view and the Bayesian
approach lies in the way quantum state preparation is understood.  In the
Bayesian view, a prepared quantum state is not determined by facts alone,
but always depends on prior beliefs, in the form of a prior assignment of
a quantum operation to a preparation device.  Facts, in the form of
measurement outcomes, are used to update the prior state (or, as in
quantum process tomography, the prior quantum operation), but they never
determine a quantum state.  By contrast, according to the
objective-preparations view, the state of a system is determined by the
preparation procedure, which can be completely specified in unambiguous,
classical terms.  The objective-preparations view holds that a quantum
state is determined by the facts about the experimental setup.  This means
that, according to the objective-preparations view, quantum states are
{\em objective}.

These considerations have an important implication for the concept of
certainty in the objective-preparations view.  Let $\ket\psi$ be a state
prepared by a preparation device, and consider the observable
$O=\ket\psi\bra\psi$, which has eigenvalues 0 and 1. If the state is
$\ket\psi$, a measurement of $O$ will give the outcome 1 with certainty.
In the objective-preparations view, this certainty is implied by the facts
about the experimental setup, independently of any observer's information
or beliefs.  Effectively, in the objective-preparations view, it is a
fact, guaranteed by the facts about the experimental setup, that the
measurement outcome will be 1. The measurement outcome is thus {\it
objectively certain}. Whatever it is that guarantees the outcome, that
guarantor is effectively an objective property.  It might be a property of
the system alone, or it might be a property of the entire experimental
setup, including the system, the preparation device, and the measurement
apparatus.  In any case, the guarantor is a property of the world external
to the agent.

The above constitutes a major problem for any interpretation that
incorporates the objective-preparations view while also maintaining that
quantum states are not part of physical reality, but are {\em
epistemic\/}~\cite{Spekkens2004}, i.e., representing information or
knowledge~\cite{Brukner2001,Mermin2002}. It is simply inconsistent to
claim that a quantum state is not part of physical reality if there are
facts that guarantee that the measurement of $O$ defined above has the
outcome 1. This is a point made by Roger Penrose in his book {\em The
Emperor's New Mind\/}, p.~340~\cite{Penrose1989}:

\begin{quote}
It is an implication of the tenets of the theory that for {\em any
state whatever}---say the state $\ket\chi$---there is a yes/no
measurement that can {\em in principle\/} be performed for which the
answer is {\bf YES} if the measured state is (proportional to)
$\ket\chi$ and {\bf NO} if it is orthogonal to $\ket\chi$. [\ldots]
This seems to have the strong implication that state-vectors must be
{\em objectively real}. Whatever the state of a physical system
happens to be---and let us call that state $\ket\chi$---there is a
measurement that can in principle be performed for which $\ket\chi$
is the {\em only\/} state (up to proportionality) for which the
measurement yields the result {\bf YES}, with {\em certainty}.
[\ldots]
\end{quote}

A more concise version~\cite{Busch-0010} of the same argument is that the
existence of this yes/no measurement ``is sufficient to warrant the
objective reality of a pure quantum state.'' The objective reality of the
quantum state follows here from the notion that there is something in the
world, independent of any agent, that guarantees the outcome YES.  This
notion in turn is implied by the objective-preparations interpretation.
The objective-preparations view is, therefore, inconsistent with the idea
that a quantum state does not represent a property of the external world.

How does our Bayesian approach escape the same conclusion? In the Bayesian
view, the quantum state of the system is not fully determined by facts
about the preparation device, since prior beliefs about the preparation
device inevitably enter into the assignment of the quantum state it
prepares.  The statement that the measurement outcome is 1 with certainty
is thus not a proposition that is true or false of the system, but an
agent's belief---and another agent might make a different prediction.
Certainty resides in the agent, not in the physical world.  In the
Bayesian approach there is no property of the system, or of the system
plus the preparation device and measurement apparatus, that guarantees
that the outcome will be 1.

\section{Certainty and locality}
\label{sec:koch}

In the preceding section, we established that the objective-preparations
view of quantum state preparation implies that there are physical
properties of the agent's external world guaranteeing measurement outcomes
occur with certainty. In this section, we show that such properties must
necessarily be nonlocal.  We show that the assumption of locality rules
out the existence of a preassigned outcome even in a measurement where the
premeasurement state is an eigenstate of the observable.  This buttresses
our previous argument that probability-1 predictions are not preordained.

Consider a measurement of the observable
\begin{equation}
O = \sum_{k=1}^d \lambda_k \ket{\phi_k}\bra{\phi_k} \;,
\end{equation}
where the states $\ket{\phi_k}$ form an orthonormal basis. If the system
state, $\ket\phi$, before the measurement is an eigenstate of $O$, say
$\ket{\phi}=\ket{\phi_j}$ for some $j\in\{1,\ldots,d\}$, the measurement
outcome will be $\lambda_j$ with probability 1. In other words, the
measurement outcome is {\em certain}. It is tempting to say, in this
situation, that $\lambda_j$ is a property that was attached to the system
already before the measurement. All the measurement would do in this case
would be to reveal this pre\"existing property of the system. This
property of the system would guarantee that the result of the measurement
will be $\lambda_j$.

We now give a version of an argument by Stairs~\cite{Stairs1983} showing
that the idea that a measurement of a system in an eigenstate of an
observable reveals a pre\"existing property of the system, or indeed a
pre\"existing property of the world external to the agent, conflicts with
locality, i.e., with the assumption that a system property at a point $x$
in space-time cannot depend on events outside the light cone centered at
$x$.

For this we consider, for the sake of concreteness, the set $S$ of 33
states in three dimensions introduced by Peres in his version of the proof
of the Kochen-Specker theorem~\cite{Peres1993a}. The 33 states in $S$ can
be completed to form 40 orthonormal bases, $\{\ket{\psi^k_1},
\ket{\psi^k_2}, \ket{\psi^k_3}\}$, $k=1,\ldots,40$, consisting of a total
of 57 distinct states~\cite{Larsson2002}.  These bases are, of course, not
disjoint.  For this set of states, one proves the Kochen-Specker theorem
by showing that there is no map,
\begin{equation}
f : S \to \{0,1\} \;,
\end{equation}
such that, for each basis $\{\ket{\psi^k_1},
\ket{\psi^k_2},\ket{\psi^k_3}\}$, exactly one vector is mapped to 1 and
the other two are mapped to 0; i.e., there are no integers
$j_k\in\{1,2,3\}$, for $k=1,\ldots,40$, such that a function on the 57
states can be defined consistently by the conditions
\begin{equation}        \label{eq:fcond}
f(\ket{\psi^k_j}) =
\left\{ \begin{array}{cc}
 1  & \mbox{if } j=j_k \;, \cr
 0  & \mbox{if } j\ne j_k \;.
\end{array} \right.
\end{equation}
We show now that the assumption of pre\"existing properties for
eigenstates, combined with the assumption of locality, implies the
existence of such an impossible map and is therefore ruled out by
locality.

Take two particles, at spatially separated locations $A$ and $B$, in the
maximally entangled state
\begin{equation}
\ket\Psi = \frac1{\sqrt3} (\ket{00} + \ket{11} + \ket{22}) \;,
\end{equation}
where the states $\ket0$, $\ket1$, $\ket2$ form an orthonormal basis. For
any single-system state
\begin{equation}
\ket\psi=c_0\ket0+c_1\ket1+c_2\ket2\;,
\end{equation}
define the complex conjugate state
\begin{equation}
\ket{\tilde\psi}=c_0^*\ket0+c_1^*\ket1+c_2^*\ket2 \;.
\end{equation}
For some $k\in\{1,\ldots,40\}$, let a von Neumann measurement in the basis
$\{\ket{\tilde\psi^k_1}, \ket{\tilde\psi^k_2},\ket{\tilde\psi^k_3}\}$ be
carried out on the particle at $A$.  Denote the outcome by
$j_k\in\{1,2,3\}$. The resulting state of the particle at $B$ is
$\ket{\psi^k_{j_k}}$.  It follows that a measurement of the observable
\begin{equation}
O^k_{j_k}=\ket{\psi^k_{j_k}}\bra{\psi^k_{j_k}}
\end{equation}
on the particle at $B$ gives the outcome 1 with certainty and that a
measurement of
\begin{equation}
O^k_{j}=\ket{\psi^k_{j}}\bra{\psi^k_{j}}\;,\quad\mbox{for $j\ne j_k$},
\end{equation}
gives the outcome 0 with certainty.  According to our assumption that such
measurements correspond to pre\"existing properties, there must be a
property of the world that guarantees the outcome 1 for the measurement of
$O^k_{j_k}$ and the outcome 0 for the two orthogonal measurements.

Locality demands that this property be independent of the measurement at
$A$.  Since the conclusions of the last paragraph hold for any choice of
measurement, $k\in\{1,\ldots,40\}$, the assumption of locality requires
that the world have physical properties that guarantee, for each state
$\ket{\psi^k_{j}}\in S$, a unique outcome $\in\{0,1\}$ for a measurement
of $\ket{\psi^k_{j}}\bra{\psi^k_{j}}$, and these properties define a map
$f : S \to \{0,1\}$ satisfying the impossible conditions (\ref{eq:fcond}).
It follows that the assumptions of locality and pre\"existing properties
for eigenstates are mutually contradictory.

Of course, one could take the position that the quantum state of
system~$B$, after the measurement on~$A$, is an objective property, but
not an objective property of $B$ alone~\cite{Grangier2005}.\footnote{N.~D.
Mermin (private communications, 2003 and 2006) characterizes as
``dangerously misleading'' the idea that the post-measurement quantum
state $|\psi\rangle$ of $B$ is an objective property of system $B$ alone.
He ``reject(s) the notion that objective properties (must) reside in
objects or have physical locations.''  Yet the quantum state of $B$, if
objective, is a property of the world, external to the agent, and since it
changes as the world changes, it is hard to see how it can have only the
disembodied objectivity Mermin is describing.  If the objectivity of
$|\psi\rangle$ {\em does\/} reside somewhere, say, in the entire
experimental setup, including the device that prepares $A$ and $B$ and the
measurement on $A$ and its outcome, then consider a measurement of an
observable of $B$ for which $|\psi\rangle$ is an eigenstate with
eigenvalue $\lambda$. If $\lambda$ is an objective property, how can it
fail to reside in $B$ (under the assumption of locality), thus making
$|\psi\rangle$ a property of $B$ after all?}  It would instead be an
objective property of the two systems, the device that prepares them, and,
more generally, of the entire history of $B$ and anything it has
interacted with.  In our view, this inevitably involves nonlocal
influences, or it leads one down the path of a many-worlds (or Everett)
interpretation.

\section{Discussion: quantum certainty}
\label{sec:cert}

The arguments in the preceding two sections imply that {\em there are no
preassigned values to quantum measurement outcomes, even outcomes that are
certain}.  In other words, there is nothing intrinsic to a quantum system,
i.e., no objectively real property of the system, that guarantees a
particular outcome of a quantum measurement.  This means that we must
abandon explanations in terms of pre\"existing properties.

The Bayesian approach sketched in the previous sections takes this
conclusion fully on board by denying objective status to any state
assignment, including pure-state assignments.  In the Bayesian approach,
there is never one unique correct quantum state assignment to a system.
Two scientists obeying all the rules of quantum mechanics can always in
principle assign different pure states to the same system, without either
of them being wrong.  The statement that an outcome is certain to occur is
always a statement relative to a scientist's (necessarily subjective)
state of belief. ``It is certain'' is a state of belief, not a
fact (see footnote~3).

A common objection to this goes as follows. Imagine a scientist who
performs a sequence of $Z$ measurements on a qubit.  Quantum mechanics,
plus his experience and prior judgment and perhaps the outcomes of a long
sequence of previous measurements, make him certain that the outcomes will
all be ``up''. Now he performs the measurements, and he always gets the
result ``up''.  Shouldn't the agent be surprised that he keeps getting the
outcome ``up''?  Doesn't this mean that it is a fact, rather than a mere
belief, that the outcomes of his experiment will be ``up''?  Doesn't this
repeated outcome demand an explanation independent of the agent's belief?

The answer to the first question is easy: Surprised?  To the contrary, he
would bet his life on it.  Since the agent was certain that he would get
the outcome ``up'' every time, he is not going to be surprised when that
happens.  Given his prior belief, only observing ``down'' would surprise
him, since he was certain this would not happen, though nature might
choose to surprise him anyway.

The answer to the second question is similarly straightforward. According
to our assumption, the agent has put together all his experience, prior
beliefs, previous measurement outcomes, his knowledge of physics and in
particular quantum theory, all to predict a run of ``up'' outcomes.  Why
would he want any further explanation?  What could be added to his belief
of certainty?  He has consulted the world {\em in every way he can\/} to
reach this belief; the world offers no further stamp of approval for his
belief beyond all the factors that he has already considered.

The third question brings us closer to a deeper motivation for this
challenge.  The question could be rephrased as follows: Isn't not asking
for a further explanation a betrayal of the very purpose of science,
namely, never to give up the quest for an explanation~\cite{Garrett1993}?
Shouldn't a naturally curious scientist never give up looking for
explanations?  The answer to these questions is that truth can be found at
different levels. At one level, a scientist who accepts the Bell/EPR
arguments should indeed stop looking for an explanation in terms of hidden
variables or preassigned values.  The Bell/EPR arguments show that there
simply is no local and realistic explanation for the correlations
predicted by quantum mechanics. Giving up the quest for such an
explanation is unavoidable if one stays within the framework of quantum
theory.  On a different level, it appears that the absence of a
mechanistic explanation is just the message that quantum mechanics is
trying to send us.  Accepting the Bell/EPR analysis at face value means
accepting what might be the most important lesson about the world, or what
we believe about the world, coming from quantum theory, namely, that there
are no instruction sets behind quantum measurement outcomes.  Go beyond
quantum mechanics if you wish to formulate an explanation in terms of
instruction sets, but accept the lesson of no instruction sets if you wish
to interpret quantum mechanics.

It might still be argued that an agent could not be certain about the
outcome ``Yes'' without an objectively real state of affairs guaranteeing
this outcome, i.e., without the existence of an underlying instruction
set.  This argument, it seems to us, is based on a prejudice.  What would
the existence of an instruction set add to the agent's beliefs about the
outcome?  Why would he be more confident about the outcome ``up'' if he
knew that the particle carried an instruction set?   The existence of
instruction sets might make the agent feel better if he is bound by a
classical world view, but from the perspective of quantum mechanics, would
not contribute to his certainty about the outcome.

Let us end with a couple of points for future research.  We have
emphasized that one of the arguments often repeated to justify that
quantum-mechanical probabilities are objective, rather than subjective, is
that they are ``determined by physical law.''  But, what can this mean?
Inevitably, what is being invoked is an idea that quantum states
$|\psi\rangle$ have an existence independent of the probabilities they
give rise to through the Born rule,
\begin{equation}
p(d)=\langle\psi| E_d | \psi\rangle\;.
\end{equation}
 From the Bayesian perspective, however, these expressions are not
independent at all, and what we have argued in this paper is that quantum
states are every bit as subjective as any Bayesian probability.  What then
is the role of the Born rule?  Can it be dispensed with completely?

It seems no, it cannot be dispensed with, even from the Bayesian
perspective.  But its significance is different than in other developments
of quantum foundations: The Born rule is not a rule for {\it setting\/}
probabilities, but rather a rule for {\it transforming\/} or {\it
relating\/} them.

For instance, take a complete set of $D+1$ observables $O^k$,
$k=1,\ldots,D+1$, for a Hilbert space of dimension $D$
~\cite{Wootters1986}.  Subjectively setting probabilities for the $D$
outcomes of each such measurement uniquely determines a quantum state
$|\psi\rangle$ (via inverting the Born rule).  Thus, as concerns
probabilities for the outcomes of any other quantum measurements, there
can be no more freedom.  All further probabilities are obtained through
linear transformations of the originals.  In this way, the role of the
Born rule can be seen as having something of the flavor of Dutch-book
coherence, but with an empirical content added on top of bare,
law-of-thought probability theory: An agent interacting with the quantum
world would be wise to adjust his probabilities for the outcomes of
various measurements to those of quantum form if he wants to avoid
devastating consequences. The role of physical law---i.e., the assumption
that the world is made of quantum mechanical stuff---is codified in how
measurement probabilities are related, not how they are
set.\footnote{These ideas mesh to some extent with Pitowsky's
development~\cite{Pitowsky2003}.  Pitowsky, however, suggests that quantum
mechanics entails a modification of probability theory, whereas we think
the Born rule is an empirical addition to probability, not a
modification.}

This brings up a final consideration.  What we have aimed for here is
to show that the subjective Bayesian view of quantum probabilities is
completely consistent, even in the case of certainty.  One of our
strong motivations for doing this is our belief that taking this
approach to quantum mechanics alleviates many of the conceptual
difficulties that have been with it since the beginning.  But even
so, this is no reason to stop digging deeper into the foundations of
quantum mechanics. For all the things the Bayesian program seems to
answer of quantum mechanics, there is still much more to question.
For instance, from our point of view, the existence of Bell
inequality violations is not particularly mysterious, but this
conceptual point does not get us much closer to a technical
understanding of the exact violations quantum mechanics does provide:
What, from a Bayesian point of view, would justify that correlations
be constrained by the Tsirelson bound~\cite{vanEnk2000-2006}? Indeed,
why is the structure of quantum probabilities (Bayesian though they
be) just the way it is?  Why does that structure find its most
convenient expression through the Hilbert-space formalism? Most
importantly, let us pose a question we never lose sight of: Given
that the Bayesian approach promises a clear distinction between the
subjective and objective, what features of the quantum formalism
beyond the ones discussed here actually correspond to objective
properties? All of these questions have no immediate answer. Yet
finding answers to them will surely lead to a better understanding of
quantum phenomena. As we see it, subjective probability is the
firmest foundation for a careful approach to that quest.

\acknowledgements

We thank N.~D. Mermin for many, many discussions over the years, for
a careful reading of an earlier draft of this paper, and for his
continued encouragement that our Bayesian points are worth {\it
trying\/} to make. (He never said we succeeded.)  We also thank three
anonymous referees, D. M. Appleby, H.~C. von Baeyer, S. J. van Enk,
J. Finkelstein, and C. G. Timpson for comments that helped improve
our exposition greatly.


\end{document}